\documentclass[12pt]{article}

\usepackage{sbc-template}
\usepackage{algorithm,algpseudocode}
\usepackage{graphicx,url,subfigure}
\usepackage{booktabs}
\usepackage{array}
\usepackage{todonotes}

\usepackage[utf8]{inputenc}

\title{Effects of Social Ties in Knowledge Diffusion:\\
case study on PLOS ONE}

\author{Felipe Eltermann\inst{1, 2}, Alan Godoy\inst{1, 3}, Fernando J. Von Zuben\inst{1} }
\address{School of Electrical and Computer Engineering -- University of Campinas\\
Campinas, São Paulo, Brazil
\nextinstitute
Elabora Consultoria\\
Campinas, São Paulo, Brazil
\nextinstitute
CPqD Foundation\\
Campinas, São Paulo, Brazil
  \email{\{elterman,godoy,vonzuben\}@dca.fee.unicamp.br}
}

\begin{document} 

\maketitle

\begin{abstract}
In order to capture the effects of social ties in knowledge diffusion, this paper examines the publication network that emerges from the collaboration of researchers, using citation information as means to estimate knowledge flow. For this purpose, we analyzed the papers published in the PLOS ONE journal finding strong evidence to support that the closer two authors are in the co-authorship network, the larger the probability that knowledge flow will occur between them. Moreover, we also found that when it comes to knowledge diffusion, strong co-authorship proximity is more determinant than geographic proximity.
\end{abstract}

Keywords: knowledge diffusion, co-authorship network, multiplex network

\section{Introduction}

It is of primary interest of organizations and society in general to promote knowledge diffusion. This interest is often twofold: first, knowledge acquisition fosters best problem solving techniques and allows better decision making; on the other hand, a knowledge producer may want to position itself as source (or holder) of a piece of knowledge. The literature highlights two knowledge-intensive activities: academic research\footnote{Public research, scientific research and academic research are commonly interchangeable and in this paper, except when explicitly put otherwise, represents research conducted in universities and governmental laboratories.} \cite{Liberman1997,Barabasi2002,Newman2004} and industrial R\&D \cite{Jaffe1993,Singh2005} and their interactions \cite{Trajtenberg1992,Cohen2002,Sorenson2006}.

Based on their capability to represent flow, exchange and communication, networks are used in literature as a functional framework for the study of organizational arrangements between economic actors \cite{Powell1990}. Considering its production-consumption structure, the study of knowledge diffusion commonly uses network-based models as a foundation \cite{Barabasi2002,Newman2004,Singh2005}.

Published documents (scientific papers and patents) are considered to be a recorded track that reflects the actual occurrence of knowledge flow between its producers and consumers \cite{Jaffe1993}. Specifically, it has been considered that knowledge flow occurs when authors cite previous works of others \cite{Jaffe1993,Singh2005,Sorenson2006} or when they collaborate as joint co-authors \cite{Sidone2016} of the same document.

In this paper, we explore the methodology used by \cite{Singh2005} to study how collaboration ties in a patent network predict the occurrence of knowledge flow between teams, applying it to a network of scientific papers. A multiplex network is considered, in which papers are represented by nodes and two layers are present: the citation layer -- in which oriented edges represent the citations between papers -- and the co-authorship layer in which edges represent the co-authorship relationships between papers.

We observed that social proximity is a strong determinant for the occurrence of knowledge diffusion in scientific research.

This paper is organized as follows: Section 2 is devoted to describing the preprocessing stages necessary to achieve to the ready-to-use dataset, followed by a brief explanation of how to built the multiplex network directly from the analysis of the dataset. Section 3 deals with the methodology adopted to extract knowledge from the multiplex network. The main results are outlined at Section 4, followed by a discussion in Section 5. Concluding remarks and further steps of the research are presented in Section 6.

\section{Methods}

The implementation of the methods described below is fully available online\footnote{https://bitbucket.org/eltermann/plosone-analysis}.
%TODO adicionar um parágrafo com um rápido overview para ajudar a ter uma visão do todo

\subsection{Citation Network and Paper-centric Co-authorship Network}

The methodology to relate social ties with knowledge flow was strongly based on \cite{Singh2005}, which argues that knowledge flow occurs when one team cites work authored by another. This definition provides a reasonable intuition for detecting the actual occurrence of knowledge flow that takes place within the scientific community. The original definition requires that, to consider a citation as corresponding to a flow of knowledge, the authors of the citing paper must be all different from the authors of the cited work. We also tested a relaxed version of this definition, considering that knowledge flow occurs provided that the citing team has at least one different author from the cited team. Our preliminary analysis showed that the difference considering both definitions was not significant, so we only present results based on the original definition.

While it is possible to map the occurrence of knowledge flow in a structure of teams or authors, greater granularity is achieved when allowing the concept of flow to be applied between two individual papers. Knowledge actually flows between people, but we consider that the concrete output of a particular occurrence of flow is embedded in the citation between two papers (provided the constraint mentioned in the previous paragraph).

Therefore, differently from the usual \emph{author-centric} co-authorship network analysis \cite{Barabasi2002,Newman2004}, it is also possible to represent the co-authorship relationship as follows: papers are nodes and there is an edge between any given pair of papers if, and only if, there is at least one person that is author of both papers. This \emph{paper-centric} definition reflects the social ties in the sense that the network evolves according to the pattern of collaboration between authors.

\begin{figure}[!ht]
  \centering
    \includegraphics[width=340pt]{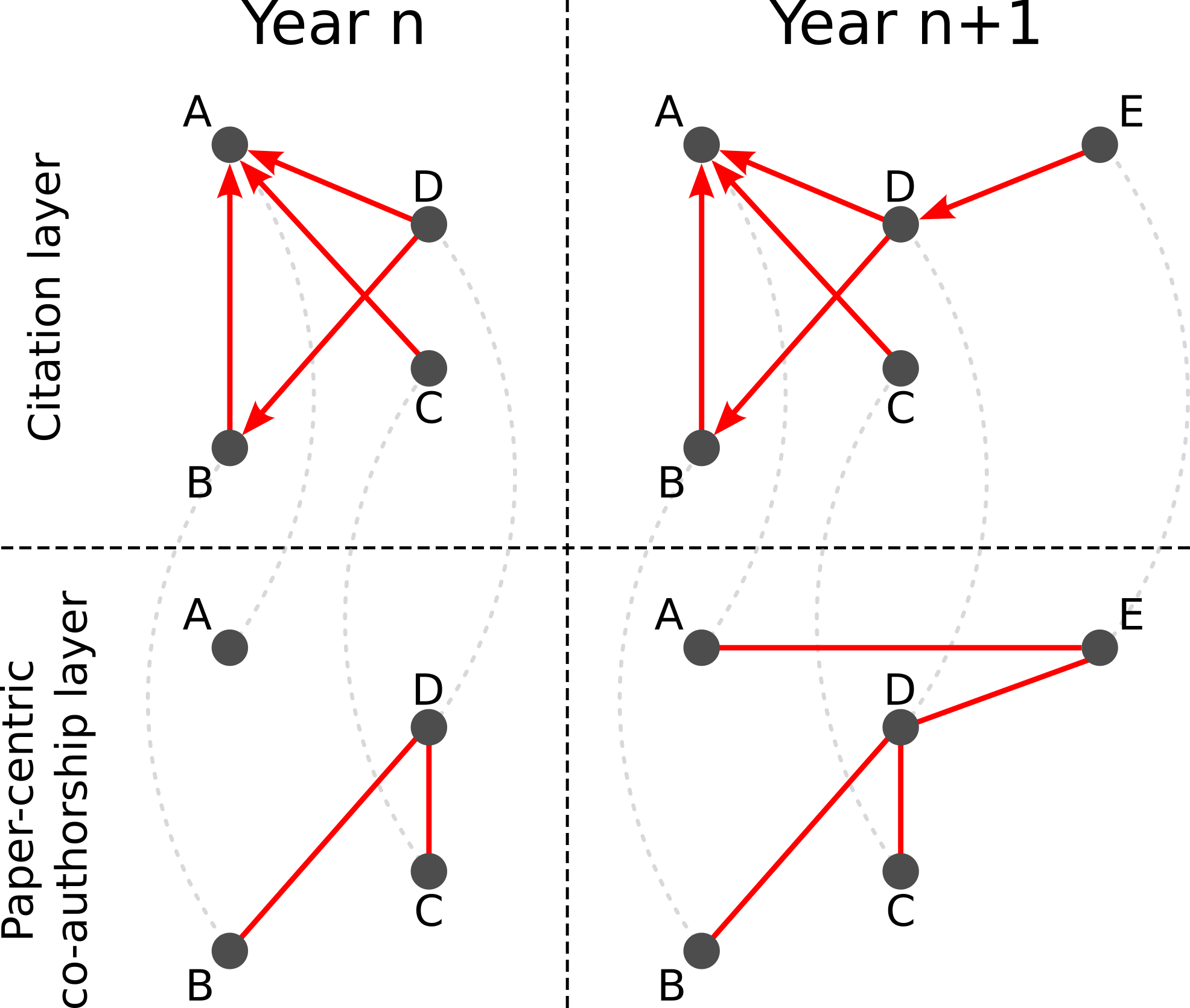}
  \caption[multiplex]{Let A, B, C, D and E be scientific papers. Also, let $a_{A}=\{1, 2\}$, $a_{B}=\{3\}$, $a_{C}=\{5\}$, $a_{D}=\{3, 4, 5\}$, and $a_{E}=\{1, 4\}$ be the sets of authors of each paper. Citations are drawn as follows: A is cited by B, C and D; B is cited by D; and D is cited by E. Finally, A, B, C and D are published in year n and E is published in year n+1. Given this setting, the top half of the figure presents the citation layer of our paper-centric network: papers are represented as nodes and there is an oriented edge from one paper to another if, and only if, the former cites the latter. The bottom half presents the co-authorship layer of the network, in which there is an edge between two papers if, and only if, they share at least one common author. The social distance between two papers is defined as the number of nodes in the shortest path between them in the co-authorship layer. Given $d_{n}(X,Y)$ as the social distance between papers X and Y in year n, we pick the following examples: $d_{n}(B,D)=0$, $d_{n+1}(B,E)=1$, $d_{n}(A,B)=\infty$, and $d_{n+1}(A,B)=2$. Now, the occurrence of knowledge flow between two papers $y(X,Y)$ is defined as a Boolean variable that is set to 1 if, and only if, one paper cites the other and there is no common author between them. Considering this definition, we have: $y(B,D)=0$, $y(B,E)=0$, and $y(A,B)=1$.}
  \label{multiplex}
\end{figure}

From the two-layered multiplex network represented in Figure \ref{multiplex}, it is possible to observe that the social distance between two papers can assume an infinite value (when the papers are in different connected components in the co-authorship layer). It is also possible to observe that the distance between two papers may vary with time, directly affecting the design of the experiment: each pair of papers must be considered in its proper period (this is detailed next).

Algorithm \ref{pseudo-code} describes how to build the paper-centric co-authorship layer of the network and to collect data of social distance and knowledge flow for later analysis. It is important to keep in mind that only the paper-centric co-authorship layer is indeed built, while information from the citation layer is implicitly used to compute the occurrence of knowledge flow.

In order to ensure that the social distances are computed correctly, the co-authorship layer expands gradually in a discrete year-by-year fashion. For each year, the graph is incremented with papers published that year and for each of those papers, pairs are formed by visiting all other nodes in the graph. By doing so, we only compute those social distances that are related to at least one paper that was published in the most recent year available in the graph. To shorten our notation, we define $d(X,Y)$ (without the subscript) as the aforementioned contextualized distance between papers X and Y. Even though it would be more precise to consider the passage of time in a more granular fashion, the use of years is imposed by the data available. This limitation is argued to be negligible because of the nature of the application: while recent papers are more likely to receive new citations \cite{Redner2005}, both research and publishing take time, imposing a delay of at least a few months between the publishing of a paper and the first appearance of its citation.

\begin{algorithm}
\caption{Extracting social distances and occurrence of knowledge flow.}
\label{pseudo-code}
\begin{algorithmic}[1]
  \State $G \gets empty\_graph$
  \State $d \gets \{\}$
  \State $y \gets \{\}$
  \For{each available year $n$ taken in order}
    \State $G.vertices \gets G.vertices \cup \{paper\ with\ year == n\}$
    \For{each paper $A$ with $year = n$}:
      \For{each paper $B$ from $G.vertices$ with $B \neq A$}:
        \If {$A$ and $B$ share at least one common author}
          \State $G.edges \gets G.edges \cup \{(A, B)\}$
        \EndIf
      \EndFor
    \EndFor
    \For{each paper $A$ with $year = n$}:
      \For{each paper $B$ from $G.vertices$ with $B \neq A$}:
        \State $d(A, B) \gets num\_intermediate\_nodes\_in\_shortest\_path(A,B)$ 
        \If {$A$ cites $B$ and there is no person being author of both $A$ and $B$}
          \State $y(A, B) \gets 1$
        \Else
          \State $y(A, B) \gets 0$
        \EndIf
      \EndFor
    \EndFor
  \EndFor
  \label{pseudo_code}
\end{algorithmic}
\end{algorithm}

\subsection{Data collection and pre-processing}

We decided to use data from PLOS ONE\footnote{\url{http://journals.plos.org/plosone}} for the following reasons: first, PLOS ONE content is published under the permissive Creative Commons license which allows its usage for any purpose\footnote{PLOS ONE also stands up for Open Access Research, a publishing format that does not charge users for having access to the full outputs of published research. For more information, see \url{https://www.plos.org/open-access}.}; second, its data is made available in XML files making it easy to be gathered and processed; third, it comprises publications from different fields providing a broad (yet limited) view on patterns of scientific publication; and finally, its large corpus of more than 150,000 papers allows the usage of tools that exploit statistical characteristics of data.

A preprocessing step is performed to transform the original XML data into a database, in which it is included: all PLOS ONE papers (and their authors), each of the cited papers (and their authors) -- which included several non-PLOS ONE papers --, and each citation made by PLOS ONE papers (information on citations made by non-PLOS ONE papers were not available). Data was collected on February 2016.

\subsection{Entities Disambiguation}
Precise connections are essential in our analysis: we want to be able to properly assign authors to papers and citations between papers. Unfortunately, no identifier is available for authors and the DOI\footnote{The digital object identifier (DOI) is a code that uniquely identifies electronic documents, as scientific papers.} was present for only a fraction of 15\% of all papers. This led to the necessity of dealing with pure text information to establish an approximate canonical representation of those entities.

When DOI is not available, matching of papers is solely based on the title information. Since it is rare that two different papers have the same title, our effort was to deal with a single paper title being spelled in slightly different ways. The method used is described as follows: non-alphabetical characters were removed, all letters were changed to uppercase and exceeding blank spaces were removed. We manually checked a random sample of one hundred papers matched by title and the result was satisfactory (at least 95\% of the matches occurred as expected).

Author names were already provided as separate parts: surname and given names. In one hand, since it is common that two different authors have similar names, it is possible to assign two different authors names to the same entity (considering the task of name matching as a classification, this case would be a false positive). On the other hand, depending on the abbreviation used, two different representations of one author's name may lead to duplicate entries in the database (that being a false negative). Our approach here was the one adopted by \cite{Barabasi2002}: surnames were changed to uppercase, and given names were represented as initials (all characters but capital letters were removed). The parts were concatenated into what we call canonical author name. We manually checked a random sample of one hundred authors matched by name and the strategy was satisfactory (at least 85\% of the matches occurred as expected).

Also, we are aware of more robust disambiguation methods that could be used. Nevertheless, we believe that the chosen method is simple enough to be easily replicated and that its false negatives introduce only a negligible noise on the effect of social ties to be analyzed.

\section{Extraction and Analysis Methodology}

The following aspects of the methodology are also implemented in the aforementioned source code we made available.

\subsection{Geographic Localization of Knowledge Flow}

Taking the geographic dimension into consideration when studying knowledge diffusion is relevant to broaden the understanding of its patterns and to provide richer information for strategic decisions of public and private institutions.

A widely accepted pattern is that knowledge flow is geographically localized: it occurs mostly between teams geographically close to each other \cite{Singh2005}. However, considering our definition of co-authorship distance (or social distance) between papers, it is reasonable to state that geographically closer papers tend to be socially closer too (for instance, in Brazil, the increase of 100 km between two researchers reduces the probability of collaboration by 16\% on average \cite{Sidone2016}). Therefore, there is space to explore the joint effects of social and geographic distances in the occurrence of knowledge flow.

Fortunately, geographic information is available in our dataset: information of country and region\footnote{Considering the hierarchy in an address line, the region is the highest entity after the country. In our dataset, it is commonly a state or a city.} is available in 99\% of PLOS ONE papers and 83\% of the others. To be precise, the country and region of each paper were inferred from the countries and regions of its authors. In order to decide which country and region to assign for each paper, we tested two approaches: taking the country and region of the first author and taking the majority of the countries and regions between the authors. The experiment revealed insignificant difference between these approaches and therefore the former one was chosen for further discussions.

When comparing a pair of papers with geographic assignments, we can then define two Boolean variables: $is\_same\_country$, which is 1 if, and only if, the two papers are from the same country; and $is\_same\_region$, which is 1 if, and only if, the two papers are from the same region. Even though these variables do not accurately represent the distance between the two addresses, it provides a good approximation.

\subsection{Variables and Representation}

Considering a pair of papers $(X,Y)$, we have so far defined three categories of variables: occurrence of knowledge flow $y(X,Y)$, co-authorship distance $d(X,Y)$, and geographic assignments $is\_same\_country(X,Y)$ and $is\_same\_region(X,Y)$. Taking into consideration that the co-authorship distances, when finite, increase much slower than the number of papers in the network \cite{Watts1998}, the distance variable was represented as a one-hot vector, in which position $i$ is 1 only if the distance between the pair of papers is equal to $i$ and all other positions are equal to 0.

In order to study the impact of each of these variables in the probability of knowledge flow, we trained Logistic Regression \cite{Bishop:2006:PRM:1162264} models and analyzed their coefficients. The higher a coefficient, the most relevant the variable is considered in determining the occurrence of knowledge flow.

\subsection{Sampling Strategy and Weighted Regression Model}

It is possible to note that the number of possible pairs of papers increases quadratically as more papers are added to the network. In fact, the straightforward approach of using all data to train regression is unnecessarily expensive, since a large portion of this data corresponds to pairs for which there is no occurrence of knowledge flow. This issue is addressed next.

In a large dataset, the occurrence of knowledge flow is a rare event, i.e. the probability of randomly choosing a pair of papers in which knowledge flow is present is very low. In other words, it was observed that the occurrence of knowledge flow ($y(X,Y)=1$) is a highly infrequent event in comparison with the volume of all pairs of papers. Therefore, an efficient method is to use all registers with $y(X,Y)=1$ and just a small sample taken from the $y(X,Y)=0$ cases:

\[num\_samples_{y(X,Y)=1} = \alpha * num\_records_{y(X,Y)=1}\]
\[num\_samples_{y(X,Y)=0} = \beta * num\_records_{y(X,Y)=0},\]

\noindent in which $num\_samples$ is the number of samples used to train the regression, $num\_records$ is the total number of samples in our database and $\alpha$ and $\beta$ are coefficients indicating the proportion of the database to be sampled.

Aiming at avoiding the introduction of bias because of this endogenous stratified sampling, we associated each sample with a weight inversely proportional to the probability of it being sampled:

\[weight_{y(X,Y)=1} = 1 / \alpha\]
\[weight_{y(X,Y)=0} = 1 / \beta\]

In other words, the unbalance between $num\_records_{y(X,Y)=1}$ and $num\_records_{y(X,Y)=0}$ is equipoised by the weighted regression. In our approach, \(\alpha\) is 1 and \(\beta\) is taken empirically: we take \(\beta\) so that $num\_samples_{y(X,Y)=1} ~ num\_samples_{y(X,Y)=0}$ and, thus, $\beta ~ number\_of\_citations / number\_of\_papers^2$.

\section{Results and Discussion}

On average, each of the 152,406 PLOS ONE papers was authored by 6.8 people and cited 43.3 papers. Given that 92\% of these papers are from the major subject Biology and Life Sciences, the high number of authors per paper is expected \cite{Newman2004}. The database resulting from the preliminary step consists of 3,373,968 papers (including papers that are cited by PLOS ONE papers), 2,435,359 authors and 6,294,099 citations.

Collected PLOS ONE papers range from 2006 to 2015. Given that computing the shortest paths in a large network is a very time-consuming task, we only managed to build the graph up to 2009, encompassing 8,340 PLOS ONE papers (we used the igraph\footnote{\url{http://igraph.org}} library and it took 7 days of computation on a DELL PowerEdge R430 server with Intel® Xeon® E5-2609 processor and 96GB of RAM). Even though this corresponds to just 5\% of the total PLOS ONE papers, we were able to observe some interesting phenomena. The resulting collaboration network consisted of 2,353,314 nodes (papers) and 313,507,251 edges.

\begin{figure}[ht!]
\centering
\subfigure[All pairs]{\includegraphics[width=160pt, height=120pt]{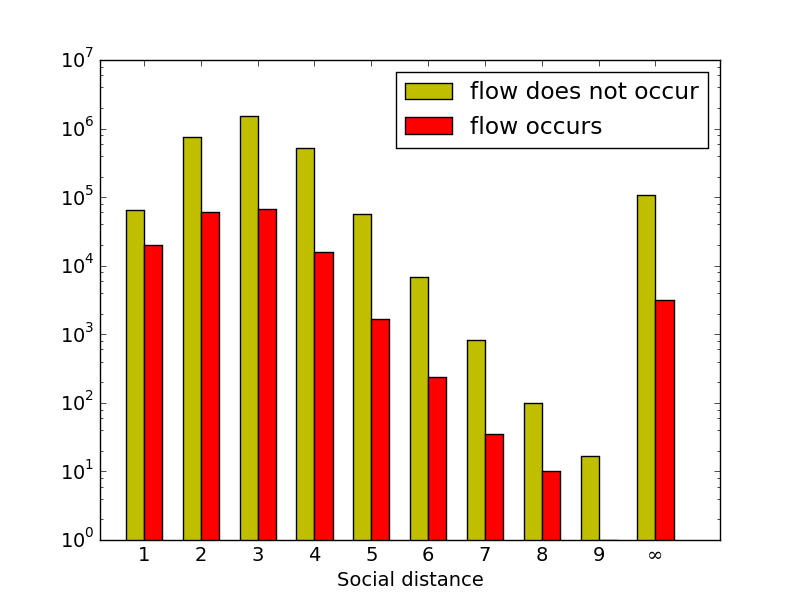}} \\
\subfigure[Same country]{\includegraphics[width=160pt, height=120pt]{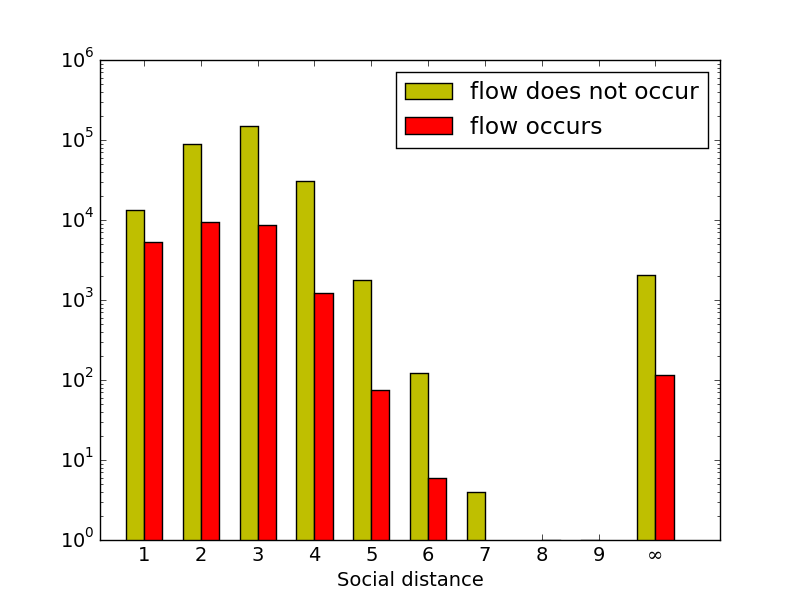}}
\qquad
\subfigure[Different countries]{\includegraphics[width=160pt, height=120pt]{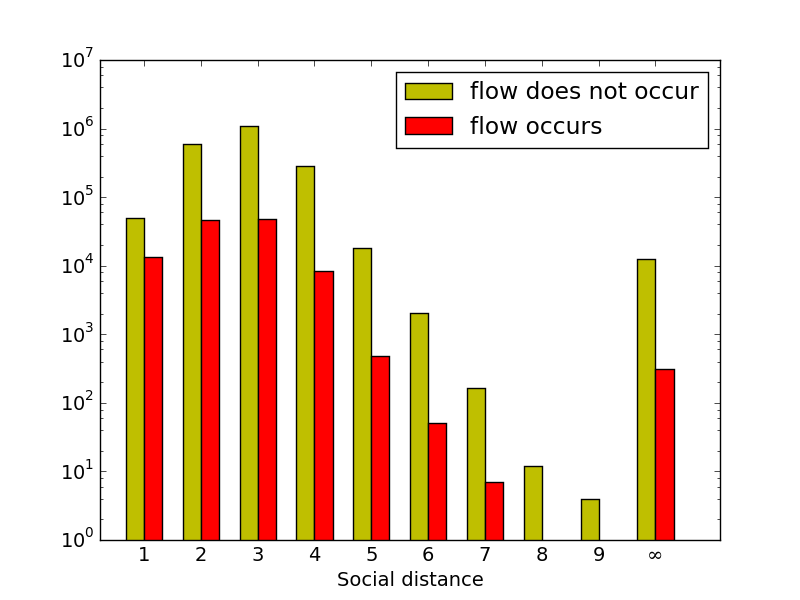}}
\qquad
\subfigure[Same region]{\includegraphics[width=160pt, height=120pt]{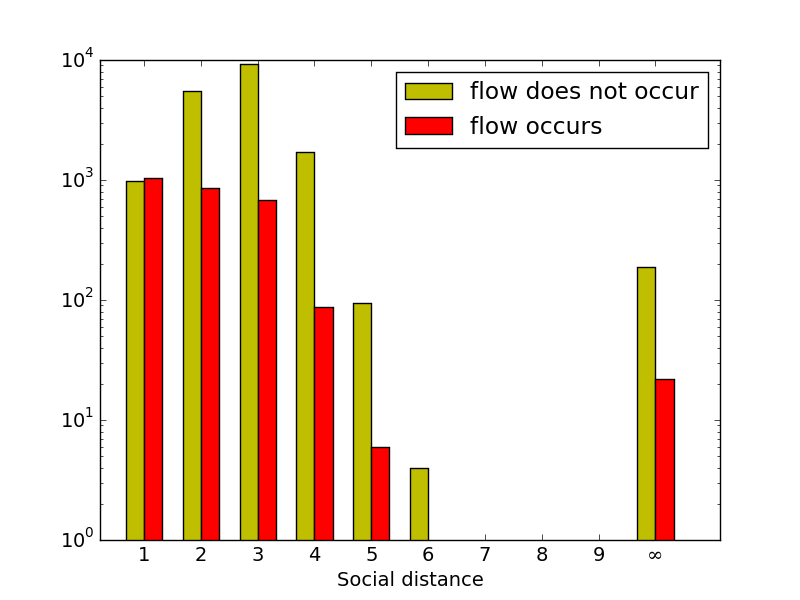}}
\qquad
\subfigure[Different regions]{\includegraphics[width=160pt, height=120pt]{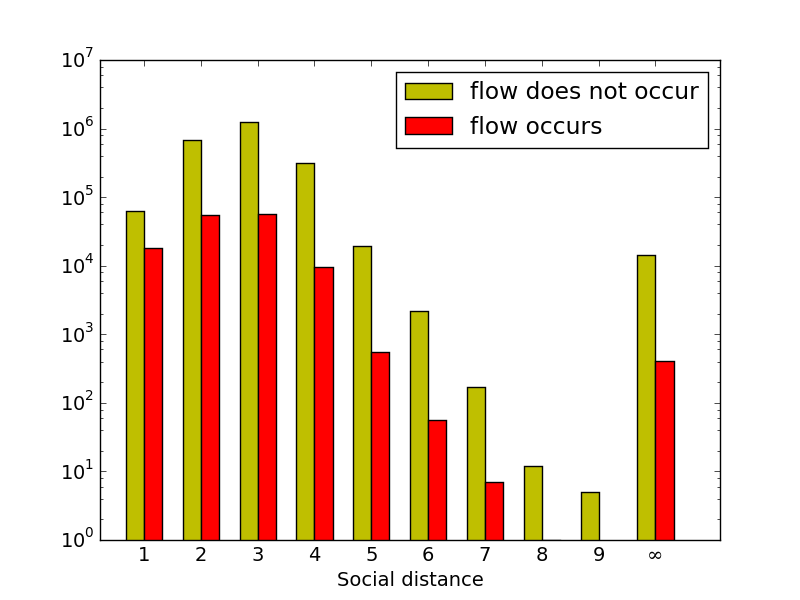}}
\caption{Histograms for 5 cohorts: (a) all sampled pairs, (b) pairs of the same country, (c) pairs of different countries, (d) pairs of the same region and (e) pairs of different regions. For each plot, the horizontal axis presents each variable of the one-hot distance vector $d(X,Y)$. The reason we do not consider the case where $d(X,Y)=0$ (both here and in the regression) is that by construction, theses cases always imply $y(X,Y)=0$. The largest observed finite social distance between two papers was 9, which reflects the small-world effect. It is also possible to note the effect of absence of geographic assignment in many papers (for instance, the sum of pairs of the same country (b) with pairs of different countries (c) is lower than the unrestricted total (a)).}
\label{results-figure-singh}
\end{figure}

\begin{figure}[ht!]
\centering
\includegraphics[width=340pt]{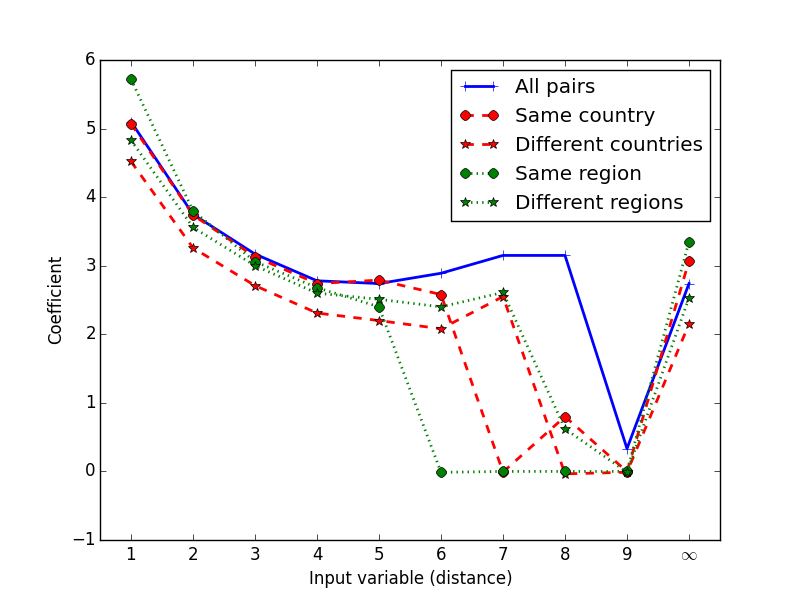}
\caption{Each set of data in Figure \ref{results-figure-singh} was used to fit the regression: the x-axis represents the binary variables of the one-hot input vectors and the y-axis corresponds to the coefficients of the regression in each case. It is possible to observe a clear decrease in relevance to predict knowledge flow as the distance increases (specially until a distance of 4 is reached).}
\label{coeffs-distances}
\end{figure}

\begin{figure}[ht!]
\centering
\includegraphics[width=340pt]{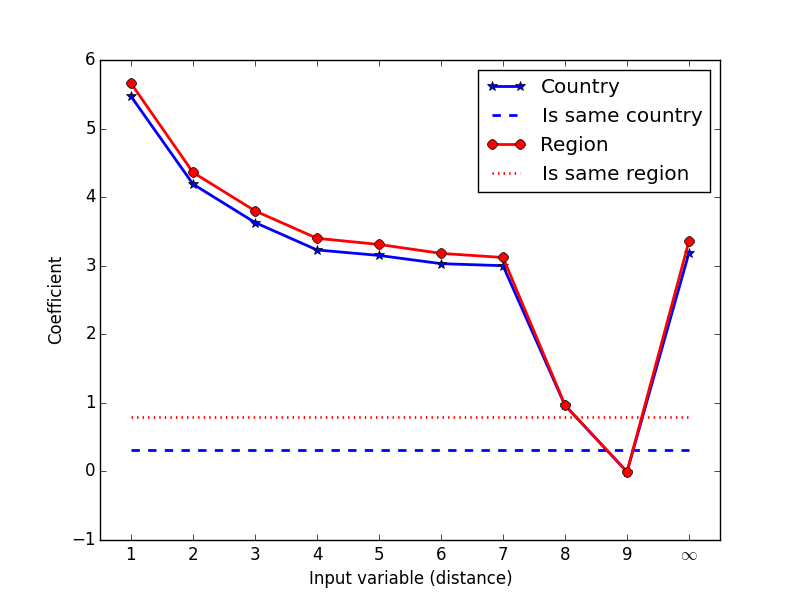}
\caption[coeffs-geography]{Data of cohorts (b) and (c) of Figure \ref{results-figure-singh} were used to generate the aforementioned Boolean variable $is\_same\_country$ and jointly fit the regression. The resulting coefficients are displayed with cross marks and the coefficient for variable $is\_same\_country$ is presented as the dashed horizontal rule. The same process was performed to data of cohorts (d) and (e) of Figure \ref{results-figure-singh} and the coefficients are presented with bullet marks. The coefficient related to variable $is\_same\_region$ is presented as the dotted horizontal rule.}
\label{coeffs-geography}
\end{figure}

The coefficients obtained for the logistic regression are displayed in Figure \ref{coeffs-distances}. They evidence that low social distances lead to a high frequency of knowledge flow, confirming the initial expectations. Despite it is possible to note a clear decrease in the effect of distance, the tendency smooths at some point. Interestingly, this point (around 3 and 4) is consonant with \cite{Christakis2013}, which states that the effects of social ties (e.g. in diffusion of behaviors) fades until a distance of approximately 3 (beyond which it becomes inconspicuous). It is important to keep in mind that this comparison was drawn between our results from a network of produced papers and others directly from social networks. A systematic exploration of the effects of an author-centric network in the corresponding document-centric network (one can name it: scientific papers, patents, software projects) would be useful to support this kind of comparison and to deepen the understanding of social collaboration.

We also fit a second set of logistic regressions, but using the binary variables $is\_same\_country$ and $is\_same\_region$ in addition to the variables indicating social distance, in order to compare the effects of geographic and social distances. Figure \ref{coeffs-geography} shows the coefficients of the regressions. It is possible to see that the coefficients related to the social distance are extremely close to those observed in the previous regressions. Also, it is noticeable that coefficients related to the geographic variables are less relevant than those related to short social distances, which indicates that being in the same geographic region matter less than being socially close to other teams for the occurrence of knowledge flow.

%Nevertheless, compared to the corresponding distance variables of the one-hot vector, the localized information of the  regional match is more useful than the broader information of the country match.
% diferença forte entra a conectividade observada aqui e aquela vista no trabalho do singh: no trabalho do Singh, mais de 40\% das patentes não pertencia ao mesmo componente conexo -- quais os possíveis motivos para isso? retrição de colaborações a serem dentro da empresa? tamanho da rede? enviesamento da amostra da PLOS ONE?

\section{Concluding Remarks}

%Even though we do not focus our attention on exploring the causation relationship between geographic proximity and social ties, the results support that despite the former is important to be taken into account when knowledge flow is important, the latter is at least of the same importance.

Comparing our results to those presented by \cite{Singh2005} related to patent data, geographic location is markedly less relevant in the present work. Such finding may indicate either that academic research is globally more integrated than industrial R\&D or that citation practices are significantly different between both communities.

Though restricted to the context of scientific production, it is possible to generalize the main consequences of the obtained results to other types of organizations. The results are suggesting that the promotion of information flow is not only influenced by the geographical proximity of key players, but also by the density of the social interactions established. Information technology is certainly a key instrument to mitigate the effects of geographical distance, together with initiatives to establish offices and representatives in strategic locations. Additionally, any organization devoted to maximizing the information flow among its peers should strongly encourage the creation of solid professional links involving external actors. This is particularly interesting for developing countries, like Brazil, as it indicates that strategic foreign social connections are a valuable path for economic and scientific growth as a result of the tapping of knowledge from abroad.

\subsection{Future work}

A clear point for improvement in this experiment would be to enrich the database. Since the information of citations was not available in any of the non-PLOS ONE papers, there is space to improve the approximation of the occurrence of knowledge flow. In that sense, co-authorship also represents only a fraction of the actual social ties between authors, and other sources of interactions could be incorporated.

Another useful analysis could be drawn considering the subject of papers, which would allow the comparison between general fields.

Finally, in order to speed-up the time for computing the distances in the co-authorship network, we plan to incorporate the usage of a GPU-based implementation to leverage the possibility of calculating the shortest paths in parallel.

\section{Acknowledgements}
We thank Elabora Consultoria for providing the computational structure to collect the data and run the experiments. We also thank Leonardo Maia for the helpful suggestions on working with data provided by PLOS ONE.

\bibliographystyle{sbc}
\bibliography{sbc-template}

\end{document}